# Studies of hot photoluminescence in plasmonically-coupled silicon via variable energy excitation and temperature dependent spectroscopy

*Carlos O. Aspetti †, Chang-Hee Cho \*\* §, Rahul Agarwal†, Ritesh Agarwal \* †*

† Department of Materials Science and Engineering, University of Pennsylvania, Philadelphia PA 19104, USA

§ DGIST-LBNL Joint Research Center & Department of Emerging Materials Science, Daegu Gyeongbuk Institute of Science and Technology (DGIST), Daegu 711-873, Korea

By coupling silicon nanowires (~150 nm diameter, 20 micron length) with an Ω-shaped plasmonic nanocavity we are able to generate broadband visible luminescence, which is induced by high-order hybrid nanocavity-surface plasmon modes. The nature of this super-bandgap emission is explored via photoluminescence spectroscopy studies performed with variable laser excitation energies (1.959 eV to 2.708 eV) and finite difference time domain simulations. Furthermore, temperature-dependent photoluminescence spectroscopy shows that the observed emission corresponds to radiative recombination of un-thermalized (hot) carriers as opposed to a Resonant Raman process.

Silicon, due to its indirect bandgap, converts excited charge carriers to heat much more readily than to light. In other words, silicon is a "dark" material in comparison to direct bandgap semiconductors, which is the main impediment to the application of Si for light emitting devices. The exceptionally low quantum yield of silicon stems from the large momentum mismatch between its conduction minima and valence band maxima.[1] To be more specific, it is this momentum mismatch which is predominantly responsible for a slow radiative recombination lifetime of milliseconds once carriers relax to the conduction band minimum (near the X-point), corresponding to a (theoretical) radiative quantum yield of $10^{-6}$.[2] The low levels of light emission may be circumvented in sub-10 nm quantum confined silicon nanocrystals,[3, 4] or nanoporous structures but this introduces significant new challenges in their integration with conventional electronic devices.[5-7] Previous work has demonstrated efficient emission in bulk silicon diodes (up to 6% at room temperature)[8] where the emission is enhanced by (1) applying a bias, which exponentially increases the equilibrium photon occupation probability,[9] (2) patterning the surface to enhance both absorption and emission,[10] and (3) using pristine float-zone silicon to suppress non-radiative scattering centers[10]. In addition to the fabrication costs of these devices, significant limitations include its restriction to band-edge emission (1.12 eV) and slow modulation rates; the recombination lifetime is still in the milliseconds range compared to nanoseconds in most direct bandgap materials.[11, 12]

Recently, Cho *et al*. demonstrated broadband, super-bandgap visible photoluminescence from non-quantum confined silicon nanowires.[13] Cho *et al*. were able to significantly enhance the spontaneous emission rate of silicon (and thereby the emission intensity) by coupling silicon to highly confined modes of a plasmonic nanocavity, using methods similar to those previously applied to cadmium sulfide, a direct bandgap material.[14] However, emission from non-thermalized carriers or "hot-luminescence" shares many characteristics to resonant-Raman spectroscopy (RRS)[15], even though hot-luminescence and RRS are fundamentally different processes; the former involves real electronic transitions whereas the latter does not. Thus, experimental verification of hot photoluminescence in silicon, and potentially other plasmonically-coupled indirect-bandgap materials, is of fundamental importance.

Silicon nanowires were integrated with plasmon nanocavities (Figures 1a, b) following a procedure similar to that previously reported by Cho et. al.[13] However, in contrast to the previous study, we used large (d ~150 nm, L ~20 micrometers) commercially obtained silicon nanowires (Sigma Aldrich). These nanowires demonstrate superior uniformity in their morphology, while their increased diameters (compared to 30 – 80 nm) result in higher order (plasmonic) cavity modes. High resolution transmission electron microscopy reveals a native oxide layer on these nanowires of 1.5-2.5 nm thickness (Figure 1c), which is used as an insulating interlayer to separate the active material from the metal and thereby prevent non-radiative recombination of charge carriers at the metal surface.[16] This layer is also used to sustain high intensity surface plasmon fields in the gap between the metal and the silicon core (Figures 1d, e). A thick silver film (300 nm) was deposited atop the silicon nanowires via electron beam or thermal evaporation. During thermal evaporation, particular care was taken to ensure a clean environment by first coating the entire chamber with a 200 nm layer of silver (base pressure of $10^{-6}$ Torr) followed by deposition on the Si nanowires, without breaking the vacuum. Silver is key for supporting surface plasmon modes in the vicinity of the silicon core, which can span the visible spectral range. Finite-difference-time-domain simulations of the silicon-oxide-silver cavity demonstrate these nanowires (150 nm diameter) are capable of sustaining high order (m >

7) hybrid surface plasmon modes which can significantly enhance spontaneous emission in silicon via the Purcell effect.[13, 14]

Optical characterization of individual nanowire samples with Ag-based plasmonic nanocavities was carried out using a home-built microscope setup equipped with a 60X, 0.7 NA objective (Nikon) that has a spatial resolution of ~600 nm. Variable-energy excitation experiments were conducted with 5 different laser lines obtained from a continuous wave argon-ion laser including: 457.9 nm, 488 nm, 496.5 nm, 501.7 nm, 514.5 nm, and a He-Ne laser (633 nm) corresponding to an energy range (1.959 eV - 2.708 eV). The incident photon flux at each wavelength was maintained constant by focusing 1 mW of incident laser power to a ~1 μm spot at all wavelengths. Photoluminescence spectra were collected using a spectrometer (Acton) coupled to a cooled CCD (charge-coupled device) with a spectral resolution of 0.1 nm. Temperature dependent measurements were conducted using a liquid nitrogen cooled cryostat (for temperatures between 77 K and room temperature). We measured plasmonically-coupled silicon nanowires with diameters ranging from 148 nm to 156 nm and at all the laser energies mentioned above.

The photoluminescence spectra of a single plasmonically-coupled silicon nanowire of diameter d = 150 nm are plotted vs. (absolute) emission energy (Figure 2a) for different laser excitation energies. The emission envelope spans the visible range and appears to have a fixed spectral width that is independent of the excitation energy. Plotting the same spectra vs. energy shift from the laser line reveals two clear high intensity subbands labeled A and B (Figure 2b), which occur at a fixed distance from the laser line, and are consistent with the previously reported data obtained with 2.708 eV excitation.[13] The extent of the emission envelope and the occurrence of high intensity bands are related to the electronic structure and phonon dispersion of silicon respectively, which are discussed below.

In indirect bandgap semiconductors, once the excited charged carrier is scattered to the electronic branch with momentum $q_e$ and relaxes along the electronic branch by scattering with phonons with momentum $q_r$, radiative recombination at the light line (with momentum q ~ 0) will require scattering with phonons of momentum $q = -(q_e + \sum q_r)$ to satisfy momentum conservation (Figure 2c).[17] It should be noted that both energy and momentum must be conserved, thus the emitted photon will have energy $E = E_{excited} - \sum E_q$, where $E_{excited}$ is the excitation energy and $\sum E_q$ is the total energy of all phonons involved in scattering. In silicon, intraband relaxation typically occurs on a picosecond timescale[18, 19] while radiative recombination has a ~10 ns lifetime near the direct bandgap[20]; thus radiative recombination is normally observed from carriers that thermalize near the minimum of the conduction band (near X point) (Figure 2c, blue curves). The spontaneous emission rate of silicon nanowires, on the other hand, may be enhanced by up to ~$10^2$-$10^3$ via the Purcell effect by coupling silicon to highly confined hybrid plasmonic-cavity modes[13] thereby making spontaneous emission competitive with the intraband relaxation process and enabling luminescence from non-thermalized states, i.e. hot-photoluminescence.

This competition between intraband relaxation and radiative recombination (Figure 2c, green curves) results in a broad emission envelope (Figure 2a) as carriers scatter back to the light line (leading to radiative recombination) but also as the carriers continue relaxing along the conduction bands. The limited number of available relaxation channels in the electronic dispersion is responsible for the apparent emission cut-off at ~2 eV. For all excitation energies, examining the electronic structure of silicon, we note the existence of multiple pathways for the relaxation of the excited carrier, i.e. towards the two local conduction band minima near the X-point (1.12 eV) and the valley at the L-point (~2.1 eV).[21] Based on momentum mismatch between the light line at k~0 and the electronic branch, the two most likely relaxation pathways (towards X and L points) are depicted in Figure 2c. A carrier excited with at high energy (> 2.1 eV) may relax towards all available conduction band minima, yet once the energy of the excited carrier is < 2.1 eV, relaxation towards the L-point will no longer be possible leading to less radiative recombination events, and thus a fixed spectral extent of the emission region. In addition, carriers excited with energy < 2.1 eV may only be absorbed to the conduction band along the X and K points (near the conduction band minima), which are also electronic states that have high-momentum mismatch from the light line (at q~0). Thus, we expect lower counts from these states due to both the lower number of emission channels and also the low availability of phonons required to scatter back to the light line from near the X point (see discussion of phonon mediated hot-luminescence below) due to the requirement of large wavevectors. We note that hot emission from states below the L-point bandgap is still possible, albeit at lower counts (Figure S2) due to the requirement of phonons with large wavevectors. Indeed, excitation at 1.959 eV yields a nearly background level spectrum (Figure 2b, yellow curve), suggesting the involvement of real electronic states for both absorption and emission. To summarize, from figure 2c it can be seen that hot-photoluminescence (like thermalized emission) is also an indirect process and requires phonon scattering to satisfy momentum conservation; however depending upon the region along the conduction band from which the carrier recombines, the required phonon momentum can be much lower in comparison to thermalized emission, thereby increasing the radiative quantum yield.

The high intensity subbands (Figure 2b) occur at a fixed energy separation from the laser line and are mediated by the phonon structure of silicon. The phonon dispersion of silicon features several flat regions in the phonon dispersion,[22] which correspond to phonons with high-density of states (high-DOS). For the experimentally observed hot luminescence range between 2.6 and 2.0 eV (band A), taking into account the electronic dispersion of silicon, the corresponding high-DOS regions of the phonon dispersion are observed near K- and L- points. The high-DOS transverse optical (TO) phonons around the K-point correspond to the energies between 56 - 64 meV, while those around the L-point correspond to 60 meV. In addition, the high-DOS transverse acoustic (TA) phonons can be found near the K- and L-points with the energies of 19 and 14 meV, respectively. Thus, the high intensity bands (A and B) could originate from the hot carrier transition assisted by the various pathways of high-DOS phonons. For example, taking into account at least 1 TO phonon to scatter to the electronic dispersion, one for intra-band relaxation, and another for scattering back to the light line, we expect a high intensity band at about 190 meV shift from the laser line (band A). Similarly, hot carrier emission involving 6 TO phonons can result in a high intensity band at ~360 meV shift (band B). However, it should be noted that the above pathways are representative cases, and the high intensity bands would be attributed to a summation of the various combinations of the high DOS phonons including TO and TA along with many other pathways involving slightly lower DOS phonons. These would be the most likely phonons to participate in radiative recombination and should result in the same high intensity bands regardless of the excitation energy ranging from 2.708 to 2.410 eV. Plotting the photoluminescence spectra of the d=150 nm plasmonically-coupled nanowire vs. energy shift from the laser line (Figure 2b), we observe broad high-intensity bands at ~190 meV and ~360 meV as expected from high-DOS phonons in silicon.

In addition, since the phonon dispersion around the high-DOS states is relatively flat, we also expect scattering from several electronic states which satisfy both momentum and energy conservation with a range of high-DOS phonons. Indeed, we observe variation of peak positions in band A by as much as 20 cm$^{-1}$ (3 meV) when examining band A of a single 150 nm nanowire at various excitation energies (Figure S2a). We observe even greater scatter in peak positions (40 cm$^{-1}$ or 5 meV) as a function of nanowire size at a single excitation energy of 2.541 eV (Figure S2b), a reflection of the fact that high-DOS phonons from a relatively broad region of the phonon dispersion may be involved in the hot-photoluminescence process. Indeed, a similar mechanism of hot photoluminescence was recently observed in organic dye molecules, where surface-plasmon enhanced spontaneous emission results in a series of peaks at fixed vibrational mode energies (also revealed by variable energy excitation) superimposed on a broadband emission envelope which is restricted due to availability of electronic states.[23]

It should be noted that the Purcell-enhanced spontaneous emission of silicon is a highly complex function of the spectral and spatial overlap between cavity modes and states that satisfy momentum and energy conservation, thereby involving three (quasi)-particles, i.e., plasmons, carriers and phonons. Noting that cavity modes for a particular nanowire are spectrally fixed, we expect there to be an excitation energy dependent modulation of the higher intensity emission bands. As discussed above, these bands occur at fixed energy shifts from the laser line (i.e. bands A and B), thus their spectral positions will change with the exciting laser and be tuned in and out of resonance with the cavity modes which are spectrally fixed for any given geometry. Both the broad emission envelope and the subbands are expected to be strongly modulated as a function of the excitation energy. In order to explore the cavity modes responsible for modulating the emission envelope, we performed finite-difference-time-domain simulations of all experimentally measured samples. The frequency domain response of the sample was obtained by averaging the Fourier transforms of the time domain fields due to all 3 orthogonal polarizations; that is two orthogonal polarizations in the plane of the nanowire cross section and a polarization along the nanowire long axis. Unlike the nanowires previously studied in the size range d < 80 nm[13], these larger nanowires with diameters d ~150 nm demonstrate markedly different cavity mode spectra characterized by higher order modes with electric field polarizations both perpendicular to the long axis (TE or transverse electric) and parallel to the long axis (TM or transverse magnetic) of the nanowire (Figure 3a). Note this convention is orthogonal to that used in some plasmonics literature where the field polarization is labeled with respect to the plane of incidence,[24] but inline with recent nanowire literature where the field polarization is referenced with respect to the nanowire long-axis.[25] Figure 3 shows the photoluminescence spectrum of the plasmonically-coupled silicon nanowire examined in Figure 2, along with the associated cavity field spectrum (Figure 3b). Note, there is a ~360 meV separation between peak 1 at 2.58 eV and peak 2 at 2.22 eV of the simulated cavity spectrum, which is responsible for the apparent "dip" in the photoluminescence spectrum. This separation is observed all across the measured size range in this work and is indicative of a transition between high order TE polarized modes and lower order TM modes with both azimuthal and radial components. As can be observed from the frequency domain electric field profiles (Figures 3c-f), these modes resemble whispering gallery modes (WGM) and will be classified as either $TE_{mn}$ for perpendicular electric field polarization or $TM_{mn}$ for parallel electric field polarization (with respect to the nanowire long-axis) and where the indices m and n correspond to the integer number of half wavelengths in the azimuthal and radial directions respectively. It should be noted that for WGM modes, the index "m" often refers to an integer number of wavelengths as circularly symmetric structures must observe the periodic boundary condition, which only allows modes at full wavelength multiples.[26] The base at the intersection of the nanowire and substrate breaks the circular symmetry and enables modes at half-wavelength multiples.[13] Following the spectra from right to left (i.e. from low energy to high energy laser excitation in Figure 3b), we observe 3 modes with perpendicular electric field polarization, which are attributed to the $TE_{71}$, $TE_{81}$, and $TE_{91}$ modes respectively. These are plasmonic modes similar to those reported by Cho et. al. (but of higher order), where the majority of the field is stored near the Si/SiO$_2$ interface.[13] The mode at ~2.6 eV (Figure 3c), on the other hand, is polarized parallel to the nanowire long-axis and has a completely different field profile, where the majority of the field is stored inside the Si core as opposed to the metal interface. This may be classified as the $TM_{11}$ mode. Referring to Figures 3c-f, we note that the field intensities (normalized to the source) within the core in both the TM and TE modes are $>10^2$. The TE modes demonstrate >1000 times intensity at the silicon surface, but this is offset by the superior spatial overlap between the active region and the cavity mode in the $TM_{11}$ mode. It has been previously demonstrated that in silicon-Au plasmonic core-shell nanowire photodetectors, both the TE and TM modes may contribute in similar magnitudes to the spectral characteristics of the system,[27] thus given the relative field intensities within the silicon core, we expect that both TE and TM modes can mediate the hot-PL process.

The question remains, as to why is there a spectral gap in emission between the TE and TM modes in the spectrum (Figure 3b). This may be understood from the inverse relationship between azimuthal mode order and quality factor that exists in plasmonic cavities. In all-dielectric WGM resonators, the quality factor scales proportional to the azimuthal order (and inversely proportional to the radial order)[28, 29], in other words this is why larger resonators, which host very high order modes, demonstrate the highest quality factors.[30] The opposite trend is true in plasmonic systems; the quality factor decreases with increasing azimuthal mode order[31]. In a surface-plasmon WGM-type resonator, increasing mode order within the same circumference implies increased confinement to the metal-dielectric interface. Metals are very lossy media[24, 32-34], thus higher interaction with the metal-interface results in increased damping of the cavity mode. By tracking quality factor of the TE cavity modes as a function of azimuthal order (Figure 3g) we observe that the quality factor drops precipitously with increasing order, which signifies that in our system mode damping becomes prohibitively high for azimuthal mode orders >9. Therefore, due to the highly damped high order TE modes we expect a decreased photoluminescence emission intensity in regions at energies higher than the highest order TE mode, but lower than the lowest order TM mode. This characteristic of the cavity modes is reflected in the emission spectra of several nanowires, which demonstrate a drop in emission intensity between bands A and B, and which was not observed in previous measurements of smaller (sub 80 nm) plasmonically-coupled nanowires.[13] For 2.708 eV excitation, we observe two broad regions of the emission envelope, one centered around 2.5 eV due to the TM mode and the other centered around 2.2 eV due to the TE modes. In fact, the features of the photoluminescence spectra due to excitation at other laser energies in the range 2.541 eV - 1.959 eV are all superimposed on the same emission envelope due to the TE cavity modes, but again, restricted in the low energy region due the silicon bandgap at the L-point (discussed above). Therefore, for larger sized Si nanowires (~ 150 nm range) we now exploit both TE and TM modes to generate hot-luminescence in larger nanowire cavities.

The size dependence (albeit a narrow range of 151 nm to 156 nm) of the emission is shown in figure 4. Photoluminescence spectra are analyzed for nanowires with sizes d=151 nm, 153 nm, and 156 nm, corresponding to Figures 4a-c respectively and excited at different laser energies. For d=151 nm, the emission spectrum excited with the 2.708 eV laser line (Figure 4a, black curve) demonstrates a dip in the emission around 2.38 eV (as in the d=150 nm sample examined above), which we attribute to cavity mode structure (Figure 4a, top). When the sample is excited with a 2.541 eV (488 nm) laser (Figure 4a, red curve) band A becomes resonant with the local minimum in the cavity field spectrum resulting in significant damping of the emission in this region. Indeed, the entire emission envelope resulting from the excitation energy of 2.541 eV and those at other laser energies up to 1.959 eV also reflects the same structure observed at 2.708 eV excitation. Figure 4e shows simulated cavity spectra for various nanowire diameters, which demonstrate a monotonic red-shift in the cavity modes with increasing diameter (as expected). As the nanowire size (although in a small range) increases (Figure 4b, c), the cavity modes redshift (Figure 4e) and the emission envelope shifts to the right (lower energies) leading to a direct modulation of subpeaks in band A (Figure 4b, red curve). The overall red shift of the emission also results in luminescence from lower energy states. Figure 4d is a magnified view of the low energy region of the spectrum (1.5 eV-1.9 eV), excited at 1.959 eV for all three nanowires. As the intensity of band A significantly decreases under excitation at 1.959 eV, the plasmonic nanowire with d = 153 nm shows a different peak spacing of ~15 meV, compared to that of ~30 meV at the other excitation energies, which differs by 1 TA phonon energy. This strongly suggests that, since the excitation energy is smaller than the energy gap at L-point (~2.1 eV), the electronic states along the <111> direction cannot contribute to the hot luminescence process, leading to a dramatic decrease in the counts and also different peak positions. Furthermore, as expected from figure 4e, increasing nanowire size leads to an increase in measured counts from low energy states due to the increased overlap between the cavity mode and the band A for the larger size nanowires.

The observed modulation of individual peaks and spectral features as a function of excitation energy is in contrast to the resonant Raman spectrum of silicon, which demonstrates little or no variation in its spectral features as a function of excitation energy in either the visible[35] or infrared frequencies.[1] To further explore the variation of the many spectral features of plasmonically-coupled silicon as a function of excitation energy, we use the photoluminescence spectra to extract the mean emission energy of the spectrum and thus the mean energy shift of the spectrum from the exciting laser energy. The mean emission energy was calculated from the photoluminescence spectra via $\bar{v} = \frac{\int v N(v) dv}{\int N(v) dv}$ where $N(v)$ is the number of measured counts at a frequency $v$.[36] The mean emission energy is then $h\bar{v}$.[37] Subtracting this value from the exciting laser energy results in a mean emission shift. Plotting the mean emission shift (of the spectrum) as a function of excitation energy (Figure 4f), we observe significant variation in the mean emission energy of individual nanowire samples (>100 meV comparing 2.410 eV excitation with 2.708 eV excitation and > 25 meV between 2.410 eV and 2.541 eV), which is a consequence of the dependence of the emission on both cavity modes and electronic structure as discussed above. Moreover, there is clear size dependence in the mean emission shift where larger wires demonstrate greater mean emission shifts and thus, stronger red shifting in the emission envelope. The red-shift of the emission envelope as a function of size is consistent with the previous discussion on size-dependent cavity modes (see above and figure 4e), where lower energy modes (for larger nanowires) enhance scattering from lower energy states. We also note that carbon contamination, and thus Raman activity of carbon, can be an issue especially when combined with silver[41]; although it is unlikely that trace amounts of carbon can yield such bright white light emission (~$10^5$ peak counts, >$10^6$ integrated counts), that the broad emission envelope is very strongly dependent on the excitation energy, and that the peaks change their positions and intensities depending on a variety of parameters. These observations are in contrast to the Raman spectrum of silicon where the mean emission energy should show negligible dependence on the exciting laser in this range. It should be noted that Surface Enhanced Raman Spectroscopy (SERS) is known to lead to a broad background, which can depend on plasmon modes.[38] Still, measurements in the range 2.410 eV-2.541 eV include all TE plasmon modes (see Figure 3b) thus; we expect any possible SERS background to result in little or no net variation of the average emission energy or shift. Furthermore, SERS spectra retain the same Raman spectral features over a broad excitation range[39], where the SERS enhancement is much more sensitive to resonance with electronic states than local field enhancement. [40]

To further test the validity of the hot-luminescence process, we examined the temperature dependence of the photoluminescence spectrum from plasmonically-coupled silicon and compared it to the known temperature dependence of other radiative and scattering processes. Raman spectroscopy typically shows a decrease in intensity with increasing temperature due to a decrease in the polarizability of a material with temperature.[42] Resonant Raman Spectroscopy of silicon also demonstrates a negative temperature dependence with increasing temperature (as the number of photons involved in electronic absorption increases with temperature, due to phonon mediated indirect absorption, thereby limiting the amount of photons involved in the Raman process). [43,44] Likewise, photoluminescence from direct-bandgap materials, such as GaAs[45] and CdS[46] also exhibit a negative temperature dependence due to increased non-radiative recombination at higher temperatures[17]. On the other hand, indirect transitions such as hot luminescence from an indirect-bandgap material (i.e. silicon) involve a competition between an increase in the non-radiative decay rate and also an increase in the radiative decay rate with temperature, as phonons are critical to mediating radiative recombination (see discussion above).

Previously, a positive temperature dependence was verified for silicon quantum dots, where increased photoluminescence was observed at higher temperatures and attributed to indirect-radiative recombination, as also confirmed by time resolved photoluminescence measurements.[47] For our plasmonically-coupled silicon samples, we measured the photoluminescence spectrum at temperatures in the range 77-300 K at 2.708 eV excitation and with a fixed laser power. We observed a monotonic increase in counts as a function of increasing temperature (Figures 5a, b), again, in contrast to the stokes-Raman spectra, which generally shows a decrease in intensity with increasing temperature. It should be noted that at higher temperatures the absorption coefficient of silicon will also increase (as it is phonon mediated), which can lead to a larger concentration of excited carriers and emitted light.[48] However, the ratio of absorption coefficients for silicon (2.708 eV) at 300 K and 77 K is 2.4, while the observed ratio of integrated counts (emission) at the same temperatures ranges from 10-20 depending on the nanowire size and laser excitation energy. Therefore, the temperature dependent change in absorption is insufficient to explain an order of magnitude increase in the measured increase in counts.  Thus, the temperature dependence of the emission intensity can be best explained by a hot-photoluminescence process, where the thermal activation of phonons that are required for intra- and inter-band relaxation can increase the radiative quantum yield.

In conclusion, we have generated bright luminescence from silicon nanowires coupled with metal nanocavities supported by high order hybrid cavity-surface plasmon modes. Photoluminescence spectroscopy at variable excitation energies reveals that silicon's electronic structure plays a key role in determining the emission intensity, while the individual sub-features of the spectrum are mediated by phonons in a hot-luminescence process. Finite difference time domain simulations elucidate the role of cavity modes in modulating the emission spectrum.

Furthermore, temperature dependent spectroscopy reveals a temperature dependence of the measured intensity that is indicative of hot-luminescence and rules out the Resonant Raman process. It should also be noted that, in addition to this experimental work, extensive theoretical work is necessary to analyze this highly complicated system featuring the interplay between phonons (bulk and interfacial for this hybrid system), plasmons, and charge carriers all of which play a role in the radiative recombination process. Finally, a direct measurement of the carrier lifetimes in plasmonically-coupled silicon would be highly desirable and is currently being pursued.

**References**

1. Liang, D.; Bowers, J. E. *Nat Photonics* **2010,** 4, (8), 511-517.

2. Pavesi, L.; Lockwood, D. J.; Alumni and Friends Memorial Book Fund., *Silicon photonics*. Springer: Berlin ; New York, 2004; p xvi, 397 p.

3. Cullis, A. G.; Canham, L. T. *Nature* **1991,** 353, (6342), 335-338.

4. Wilson, W. L.; Szajowski, P. F.; Brus, L. E. *Science* **1993,** 262, (5137), 1242-1244.

5. Hirschman, K. D.; Tsybeskov, L.; Duttagupta, S. P.; Fauchet, P. M. *Nature* **1996,** 384, (6607), 338-341.

6. Walters, R. J.; Bourianoff, G. I.; Atwater, H. A. *Nat Mater* **2005,** 4, (2), 143-146.

7. Park, N. M.; Kim, T. S.; Park, S. J. *Appl Phys Lett* **2001,** 78, (17), 2575-2577.

8. Trupke, T.; Zhao, J. H.; Wang, A. H.; Corkish, R.; Green, M. A. *Appl Phys Lett* **2003,** 82, (18), 2996-2998.

9. Wurfel, P. *J Phys C Solid State* **1982,** 15, (18), 3967-3985.

10. Green, M. A.; Zhao, J. H.; Wang, A. H.; Reece, P. J.; Gal, M. *Nature* **2001,** 412, (6849), 805-808.

11. Matsusue, T.; Sakaki, H. *Appl Phys Lett* **1987,** 50, (20), 1429-1431.

12. Wiesner, P.; Heim, U. *Phys Rev B* **1975,** 11, (8), 3071-3077.

13. Cho, C.-H.; Aspetti, C. O.; Park, J.; Agarwal, R. *Nat Photon* **2013,** 7, (4), 285-289.

14. Cho, C.-H.; Aspetti, C. O.; Turk, M. E.; Kikkawa, J. M.; Nam, S.-W.; Agarwal, R. *Nat Mater* **2011,** 10, (9), 669-675.

15. Permogorov, S. *Phys Status Solidi B* **1975,** 68, (1), 9-42.

16. Anger, P.; Bharadwaj, P.; Novotny, L. *Physical Review Letters* **2006,** 96, (11), 113002.

17. Pankove, J. I., *Optical processes in semiconductors*. Prentice-Hall: Englewood Cliffs, N.J.,, 1971; p xvii, 422 p.

18. Goldman, J. R.; Prybyla, J. A. *Physical Review Letters* **1994,** 72, (9), 1364-1367.

19. Sabbah, A. J.; Riffe, D. M. *Phys Rev B* **2002,** 66, (16).

20. Prokofiev, A. A.; Moskalenko, A. S.; Yassievich, I. N.; de Boer, W. D. A. M.; Timmerman, D.; Zhang, H.; Buma, W. J.; Gregorkiewicz, T. *Jetp Lett+* **2009,** 90, (12), 758-762.

21. Chelikowsky, J. R.; Cohen, M. L. *Phys Rev B* **1974,** 10, (12), 5095-5107.

22. Wei, S. Q.; Chou, M. Y. *Phys Rev B* **1994,** 50, (4), 2221-2226.

23. Itoh, T.; Yamamoto, Y. S.; Tamaru, H.; Biju, V.; Murase, N.; Ozaki, Y. *Phys Rev B* **2013,** 87, (23).

24. Maier, S. A., *Plasmonics: Fundamentals and Applications*. Springer: New York, New York, 2007.

25. Cao, L. Y.; White, J. S.; Park, J. S.; Schuller, J. A.; Clemens, B. M.; Brongersma, M. L. *Nat Mater* **2009,** 8, (8), 643-647.

26. Nobis, T.; Kaidashev, E. M.; Rahm, A.; Lorenz, M.; Grundmann, M. *Physical Review Letters* **2004,** 93, (10).

27. Fan, P. Y.; Chettiar, U. K.; Cao, L. Y.; Afshinmanesh, F.; Engheta, N.; Brongersma, M. L. *Nat Photonics* **2012,** 6, (6), 380-385.

28. Oraevsky, A. N. *Quantum Electron+* **2002,** 32, (5), 377-400.

29. Matsko, A. B.; Ilchenko, V. S. *Ieee J Sel Top Quant* **2006,** 12, (1), 3-14.

30. Vernooy, D. W.; Ilchenko, V. S.; Mabuchi, H.; Streed, E. W.; Kimble, H. J. *Opt Lett* **1998,** 23, (4), 247-249.

31. Min, B. K.; Ostby, E.; Sorger, V.; Ulin-Avila, E.; Yang, L.; Zhang, X.; Vahala, K. *Nature* **2009,** 457, (7228), 455-U3.

32. Oulton, R. F.; Sorger, V. J.; Genov, D. A.; Pile, D. F. P.; Zhang, X. *Nat Photon* **2008,** 2, (8), 496-500.

33. Khurgin, J. B.; Sun, G. *Appl Phys Lett* **2012,** 100, (1).

34. Oulton, R. F. *Nat Photonics* **2012,** 6, (4), 219-221.

35. Piscanec, S.; Ferrari, A. C.; Cantoro, M.; Hofmann, S.; Zapien, J. A.; Lifshitz, Y.; Lee, S. T.; Robertson, J. *Mat Sci Eng C-Bio S* **2003,** 23, (6-8), 931-934.

36. Epstein, R. I.; Sheik-Bahae, M., *Optical refrigeration : science and applications of laser cooling of solids*. Wiley-VCH: Weinheim, 2009; p xv, 241 p.

37. Zhang, J.; Li, D. H.; Chen, R. J.; Xiong, Q. H. *Nature* **2013,** 493, (7433), 504-508.

38. Mahajan, S.; Cole, R. M.; Speed, J. D.; Pelfrey, S. H.; Russell, A. E.; Bartlett, P. N.; Barnett, S. M.; Baumberg, J. J. *J Phys Chem C* **2010,** 114, (16), 7242-7250.

39. Jung, Y. M.; Sato, H.; Ikeda, T.; Tashiro, H.; Ozaki, Y. *Spectrochim Acta A* **2004,** 60, (8-9), 1941-1945.

40. Yoon, I.; Kang, T.; Choi, W.; Kim, J.; Yoo, Y.; Joo, S. W.; Park, Q. H.; Ihee, H.; Kim, B. *J Am Chem Soc* **2009,** 131, (2), 758-762.

41. Otto, A. *Surf Sci* **1978,** 75, (2), L392-L396.

42. Venkateswarlu, K. *Nature* **1947,** 159, (4029), 96-97.

43. Compaan, A.; Trodahl, H. J. *Phys Rev B* **1984,** 29, (2), 793-801.

44. Compaan, A.; Lee, M. C.; Trott, G. J. *Phys Rev B* **1985,** 32, (10), 6731-6741.

45. Jiang, D. S.; Jung, H.; Ploog, K. *J Appl Phys* **1988,** 64, (3), 1371-1377.

46. Hoang, T. B.; Titova, L. V.; Jackson, H. E.; Smith, L. M.; Yarrison-Rice, J. M.; Lensch, J. L.; Lauhon, L. J. *Appl Phys Lett* **2006,** 89, (12).

47. Kwack, H. S.; Sun, Y.; Cho, Y. H.; Park, N. M.; Park, S. J. *Appl Phys Lett* **2003,** 83, (14), 2901-2903.

48. Weakliem, H. A.; Redfield, D. *J Appl Phys* **1979,** 50, (3), 1491-1493.

**Supporting Information**. Additional discussion and figures. This material is available free of charge via the Internet at:

AUTHOR INFORMATION

**Corresponding Authors**

* (R.A.) Corresponding author. E-mail: riteshag@seas.upenn.edu

** (C.H.C.) Corresponding author. Email: chcho@dgist.ac.kr

**Author Contributions**

The manuscript was written through contributions of all authors. All authors have given approval to the final version of the manuscript.

**Acknowledgement**

This work was supported by the U.S. Army Research Office under Grant No. W911NF-09-1-0477 and W911NF-11-1-0024, and the National Institutes of Health through the NIH Director's New Innovator Award Program, 1-DP2-7251-01. C. O. A. is supported by the United States Department of Defense, Air Force Office of Scientific Research, National Defense Science and Engineering Graduate (NDSEG) Fellowship. C.H.C acknowledges the Leading Foreign Research Institute Recruitment Program (Grant No. 2012K1A4A3053565) and the DGIST R&D Program (13-BD-0401) through the National Research Foundation of Korea (NRF) funded by the Ministry of Science, ICT and Future Planning.

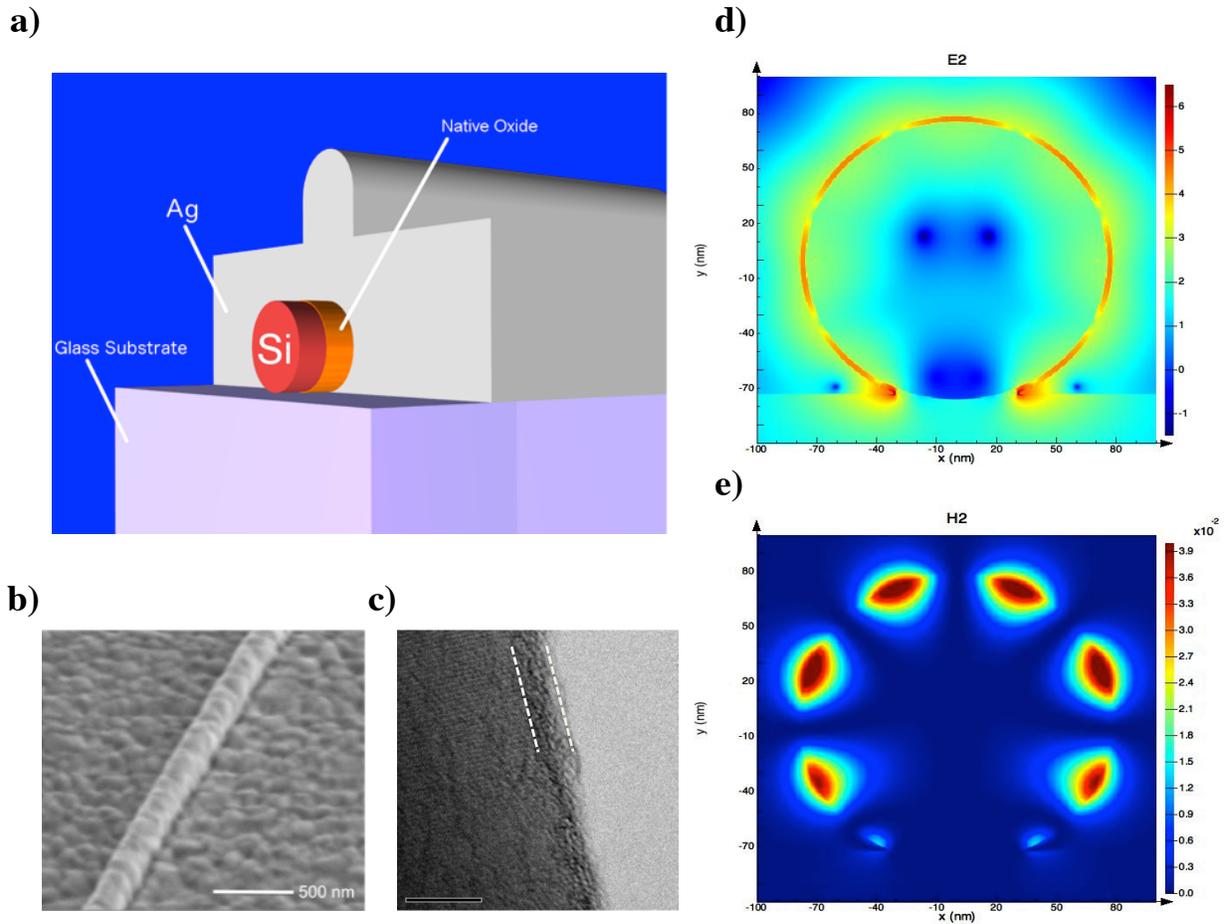

**Figure 1.** (a) Schematic of silicon nanowire integrated with a 300 nm thick silver film to form a plasmonic nanocavity (drawn to scale). The native oxide of silicon ($SiO_x$) is used to separate the active silicon core from the silver shell. (b) Scanning electron microscope (SEM) of silver coated silicon nanowire. (c) Transmission electron microscope (TEM) image of a representative bare silicon nanowire demonstrating 1.5-2.5 nm of native oxide (denoted by dashed white lines) on the nanowire surface. (d) Frequency domain spatial distribution of the electric field intensity in Ω-cavity Si (d = 150 nm) demonstrating high order (m=9) mode (obtained via FDTD simulation) and (e) corresponding magnetic field intensity.

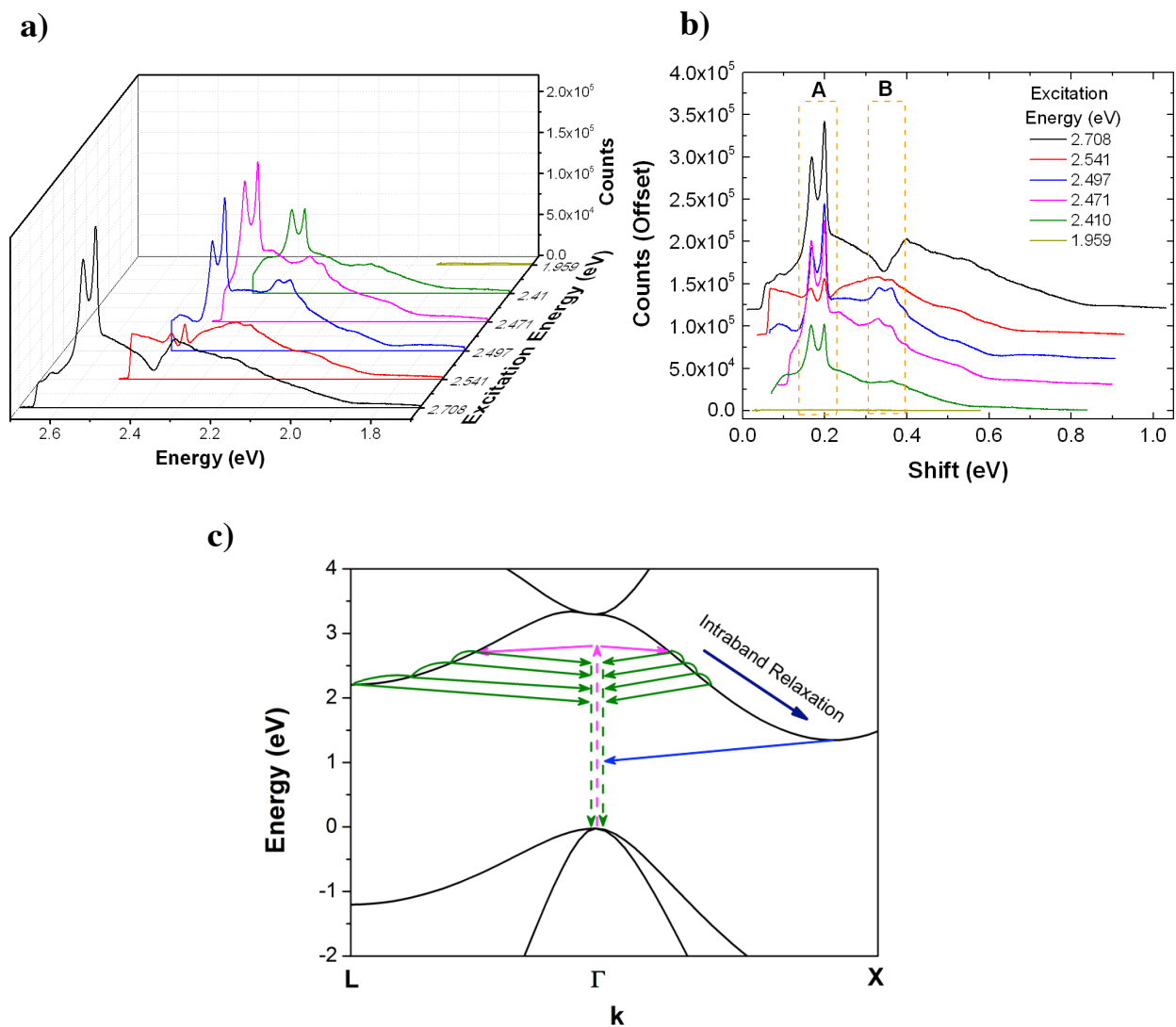

**Figure 2.** Photoluminescence spectrum of plasmonically-coupled silicon nanowire (d=150 nm) plotted vs. (a) absolute emission energy and (b) energy shift from the laser line for various excitation energies in the range 1.959 eV-2.708 eV. Spectra in (b) are plotted with a constant offset of $2\times10^3$ counts for clarity. The dashed boxes labeled A and B denote high-intensity emission bands. (c) Schematic of the electronic dispersion of silicon featuring carrier excitation (magenta arrows), intra-band relaxation and hot-luminescence (green arrows) and radiative recombination from thermalized carriers (blue arrows). This process is examined for relaxation towards the selected conduction band minima at both the X-points (1.12 eV) and L-points (~2.1 eV).

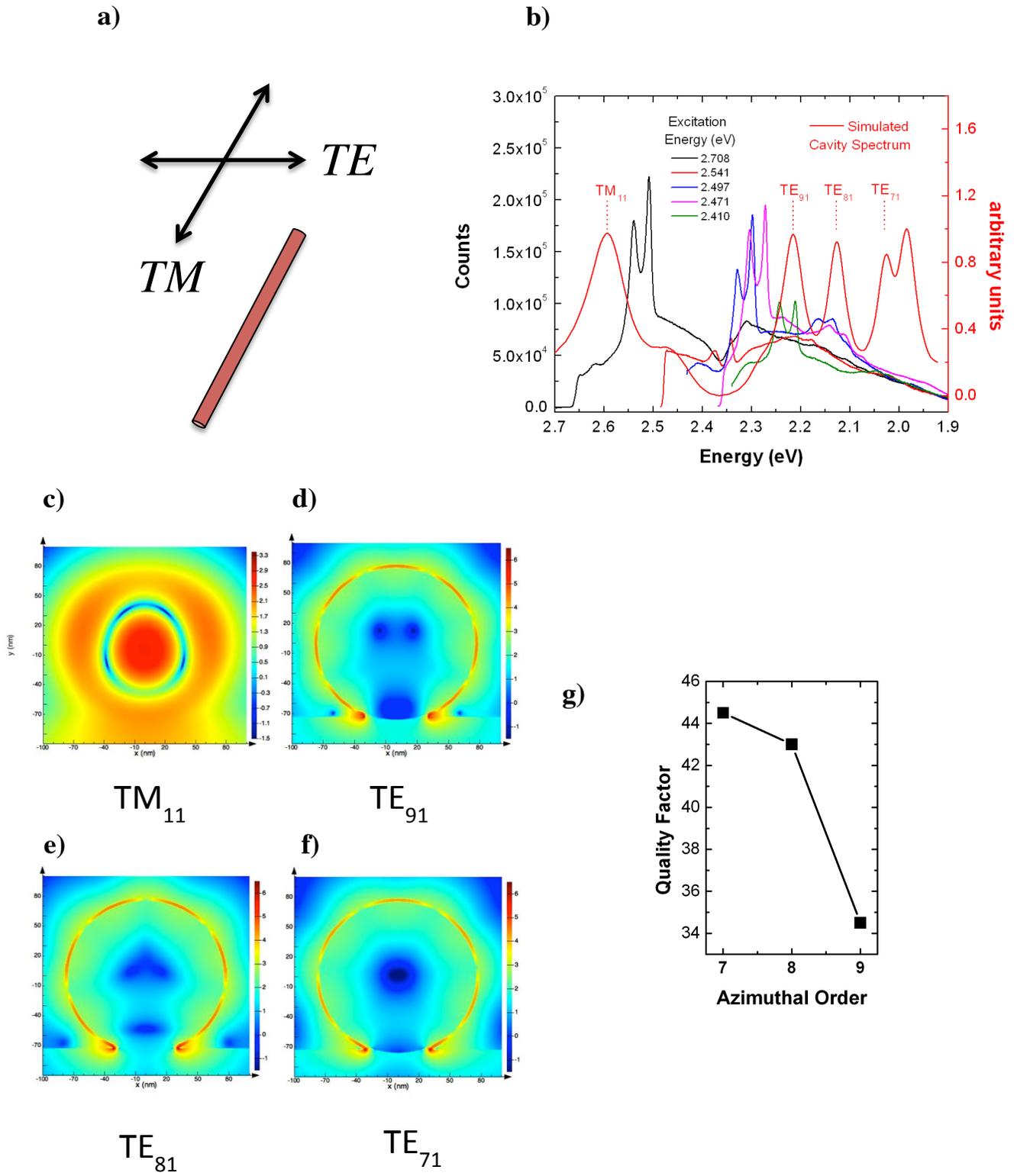

**Figure 3**

**Figure 3.** Electromagnetic mode properties of plasmonically-coupled silicon analyzed via FDTD simulations and photoluminescence spectroscopy. (a) nomenclature convention for modes polarized parallel (TM) and perpendicular (TE) to the nanowire long-axis. (b) Variable-energy excitation photoluminescence spectra of d=150 nm Ω-cavity silicon nanowire juxtaposed with simulated cavity mode spectrum (red curve). (c)-(f) frequency domain profiles of the electric intensity (log scale) for cavity modes ordered from highest to lowest energy. (h) Plot of quality factor versus azimuthal index (m), for TE modes in (b) and represented by the field profiles in (d)-(f).

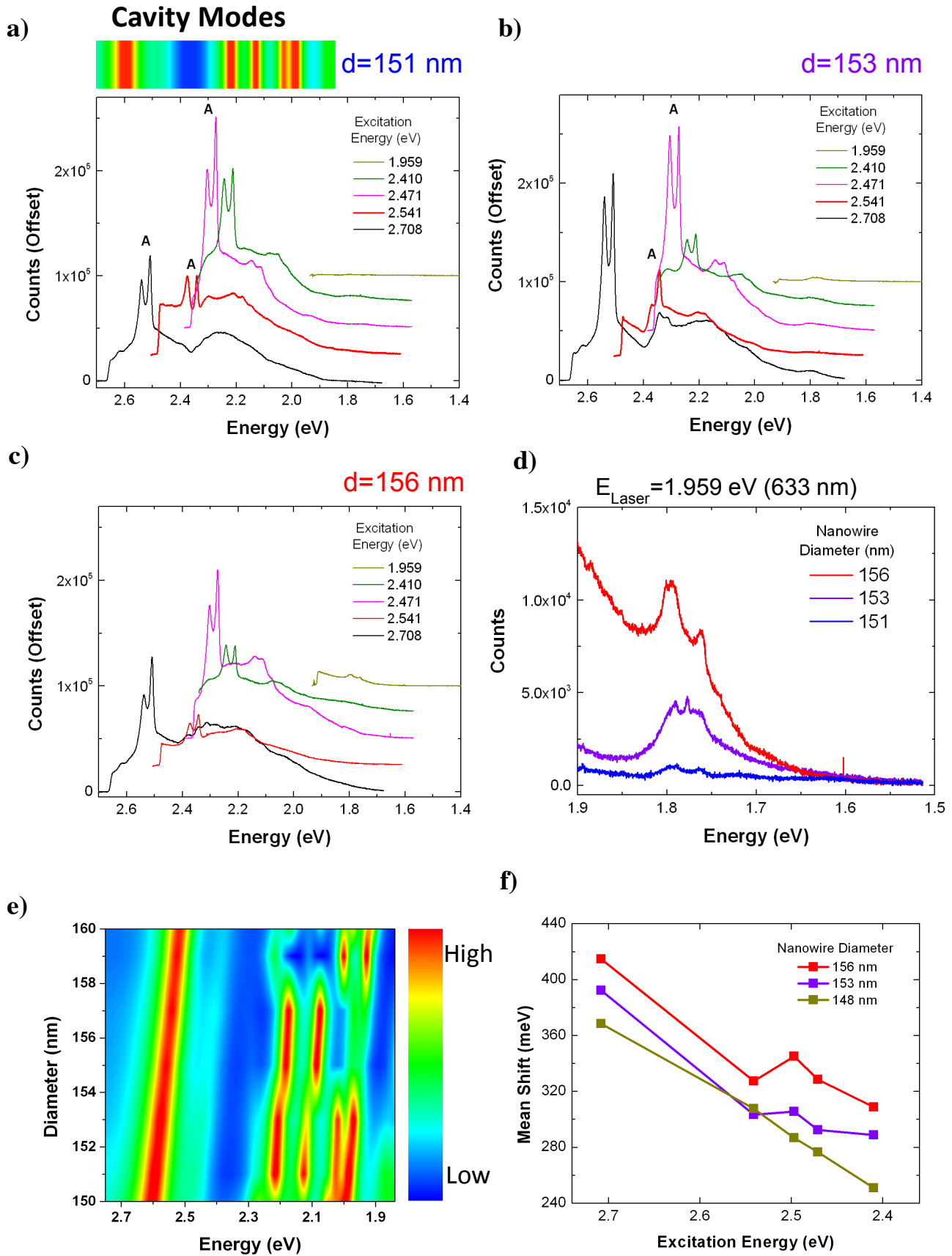

**Figure 4**

**Figure 4.** (a)-(c) Photoluminescence spectra of plasmonically-coupled nanowires excited at various laser energies in the range 1.959 eV – 2.708 eV for nanowires of size (a) d=151 nm, (b) d=153 nm, and (c) d=156 nm (all spectra are offset by 25,000 counts for clarity). The cavity mode spectrum of the d=151 nm nanowire is plotted on top of the photoluminescence spectra (high in red to low in blue) using the same energy scale. The variable energy excitation photoluminescence spectra demonstrate the role of mode structure in modulating high intensity sub-peaks. In addition to size-dependent peak modulation, the modes, which red-shift with increasing size, also enable hot luminescence at lower energies for larger nanowires. (d) Photoluminescence spectrum in low energy region (excited with 633 nm, He-Ne laser) for samples (a)-(c). (e) Simulated cavity mode spectra of plasmonically-coupled silicon nanowires with diameters in the range d=150 nm to 160 nm as a function of energy. (f) Difference between the average emission energy and exciting laser energy (i.e. the mean emission shift) plotted against excitation energy. Large fluctuations in mean emission intensity as function of size and excitation energy highlight the role of both cavity modes and electronic structure in modulating the emission spectrum.

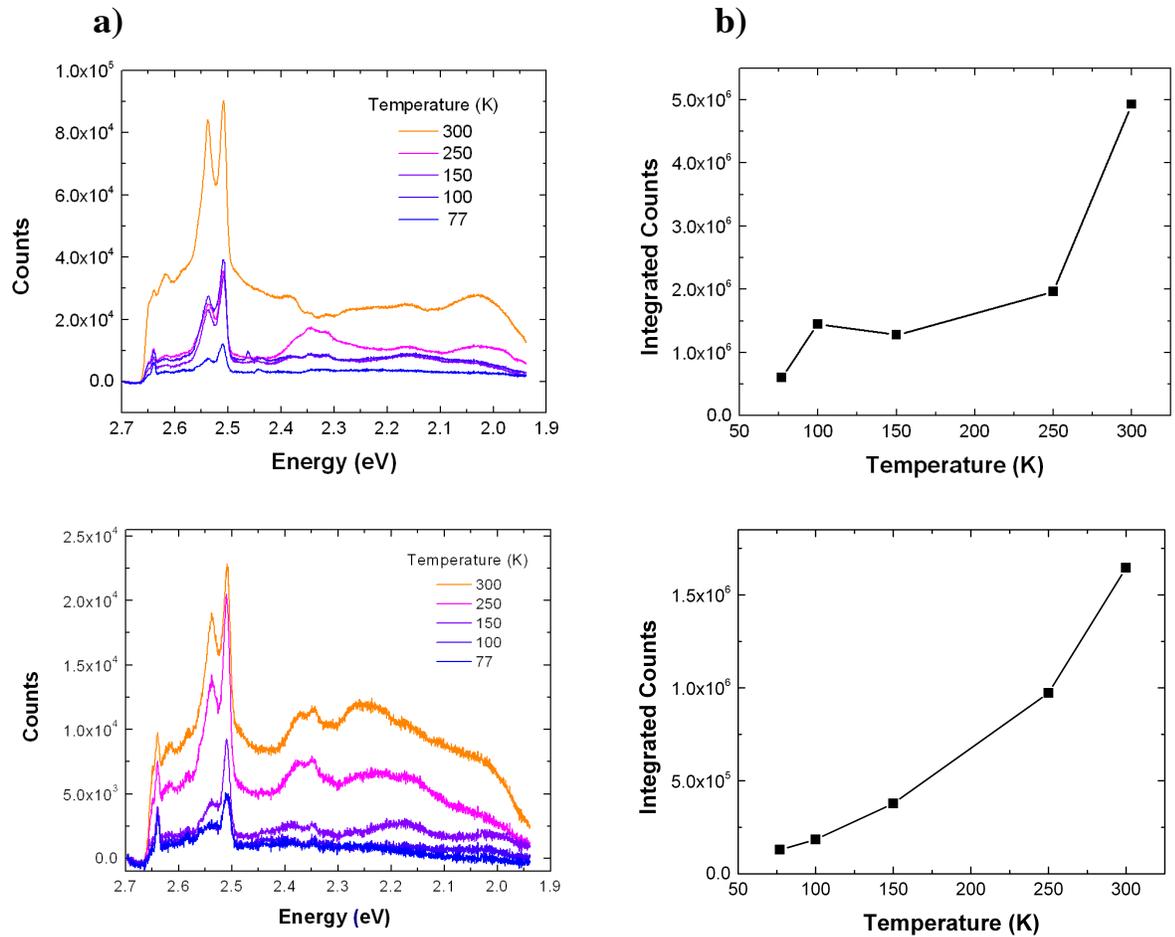

**Figure 5.** (a) Temperature dependent photoluminescence spectra of plasmonically-coupled silicon (for two different samples) in the range 77K-300K. Increase in overall emission intensity with temperature follows expected trend for hot-luminescence from an indirect bandgap material as the phonon population increases with increasing temperature. (b) Plot of total integrated counts as a function of temperature for samples shown in (a).

**Supplementary Materials for "Studies of hot photoluminescence in plasmonically-coupled silicon via variable energy excitation and temperature dependent spectroscopy"**

*Carlos O. Aspetti †, Chang-Hee Cho ** §, Rahul Agarwal †, Ritesh Agarwal * †*

† Department of Materials Science and Engineering, University of Pennsylvania, Philadelphia PA 19104, USA

§ DGIST-LBNL Joint Research Center & Department of Emerging Materials Science, Daegu Gyeongbuk Institute of Science and Technology (DGIST), Daegu 711-873, Korea

**Corresponding Authors**

* (R.A.) Corresponding author. E-mail: riteshag@seas.upenn.edu

** (C.H.C.) Corresponding author. Email: chcho@dgist.ac.kr

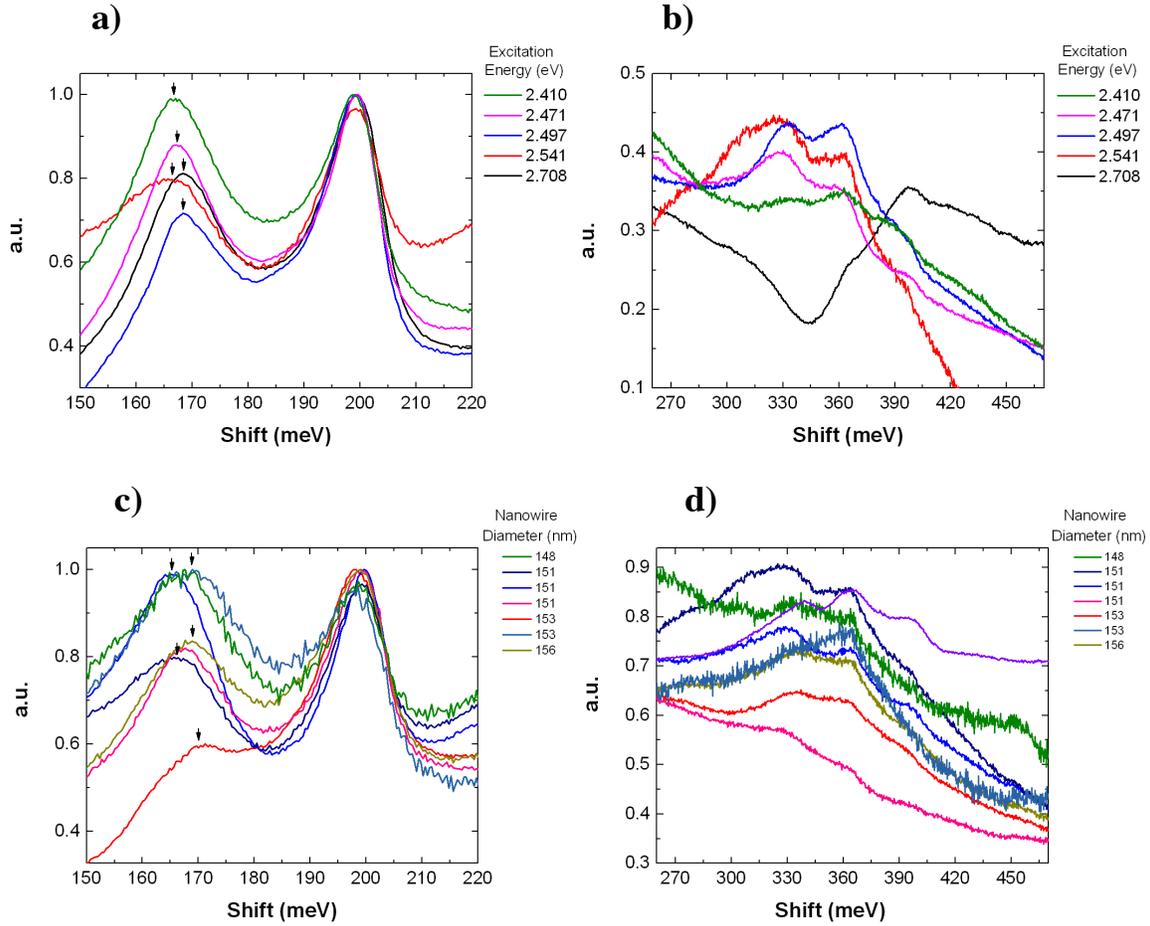

**Figure S1.** Variation in spectral positions of subpeaks in photoluminescence spectrum of plasmonically-coupled silicon nanowire. (a) variation in band A and (b) band B as a function of excitation energy for a single silicon nanowire size (d=150 nm). (c) variation in spectral positions of band A and (d) band B for several nanowire sizes as a function of excitation energy. Black arrows denotes location of peak 1.

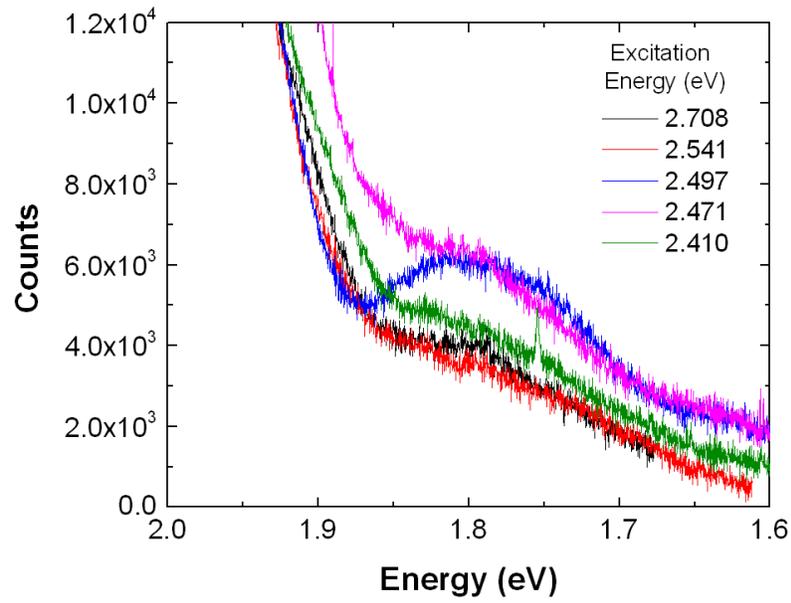

**Figure S2.** Magnified photoluminescence spectra of d=150 nm plasmonically-coupled silicon nanowire in low energy region demonstrating emission below the silicon bandgap at the L-point.